\documentstyle[pre,aps,psfig]{revtex}

\begin{document}%
%%%%%%%%%%%%%%%%%%%%%%% LANL title page start %%%%%%%%%%%%%%%%%%%%%%
%%  {\pagestyle{empty}%
%%  %
%%  \parindent0pt\sf
%%  %
%%  LA-UR submitted\\~\\
%%  %Approved for public release;\\distribution is unlimited
%%  
%%  \vfill
%%  
%%  \begin{center}
%%  
%%  {\Huge\sf A Simplified Cellular Automaton Model for City Traffic}
%%  
%%  \vfill
%%  
%%  {\large\sf DRAFT \today, comments welcome}
%%  
%%  \vfill
%%  
%%  {\large\sf Authors: Patrice Simon and Kai Nagel}
%%  
%%  \end{center}
%%  
%%  \vfill
%%  \vfill
%%  
%%  {\Huge\sf LOS ALAMOS}
%%  
%%  {\large\sf NATIONAL LABORATORY}
%%  
%%  Los Alamos National Laboratory, an affirmative action/equal
%%  opportunity employer, is operated by the University of California for
%%  the U.S. Department of Energy under contract W-7405-ENG-36.  By
%%  acceptance of this article, the publisher recognizes that the U.S.\
%%  Government retains a non-exclusive, royalty-free license to publish or
%%  reproduce the published form of this contribution, or to allow others
%%  to do so, for U.S.\ Government purposes.  Los Alamos National
%%  Laboratory requests that the publisher identify this article as work
%%  performed under the auspices of the U.S. Department of Energy.  The
%%  Los Alamos National Laboratory strongly supports academic freedom and
%%  a researcher's right to publish; as an institution, however, the
%%  Laboratory does not endorse the viewpoint of a publication or
%%  guarantee its technical correctness. 
%%  }
%%  \eject
%%  
%%  ~\vfill\eject
%%  \setcounter{page}{1}
%%%%%%%%%%%%%%%%%%%%%% LANL title page end %%%%%%%%%%%%%%%%%%%%%%
\title{\bf A Simplified Cellular Automaton Model for City Traffic\\
~\\
DRAFT \today}
\author{ P. M. Simon $^{\star\dag}$ and K. Nagel $^{\star\dag}$\\}
\address{
Los Alamos National Laboratory, TSA-DO/SA, MS-M997, Los Alamos, NM,
87545, USA $^{\dag}$,\\
Santa Fe Institute, 1399 Hyde Park Rd, Santa Fe NM 87501, USA $^{\star}$.
{\tt (simonp,kai)@lanl.gov}}

\twocolumn

\maketitle

\begin{abstract}%
We systematically investigate the effect of blockage sites in a
cellular automaton model for traffic flow.  Different scheduling
schemes for the blockage sites are considered.  None of them returns a
linear relationship between the fraction of ``green'' time and the
throughput.  We use this information for a fast implementation of
traffic in Dallas.
\end{abstract}

\section{Introduction}

In today's crowded world, space and money to build transportation
systems which can fulfill all demand is often not available, or it is
not desired to spend it on transportation system infrastructure.  The
result is congestion: from congested urban centers to congested
inner-city corridors to congested railways and congested airports.  In
consequence, some ``forecasting'' tool would be desirable.
Unfortunately, congestion has the side effect that causal relations
become much more spread both spatially and
temporally~\cite{Nagel:Rasmussen}.  If a road is crowded, the person
may attempt a different route or a different mode (spatial spreading),
or she may attempt the trip at a different time (temporal spreading)
or even totally drop the trip.  The result is that planning tools need
to consider a much wider spatial and temporal context than ever
before.  Conceptually this means that for such problems the method
needs to be ``activity based'', i.e.\ one needs to consider the whole
process how people plan transportation in a daily or better weekly
context (see, e.g.,~\cite{TRANSIMS}).

Another effect of being in the congested regime is that one needs to
worry a lot more than before about having a {\em dynamically\/}
correct representation of the transportation system: For example, a
peak-period spreading of traffic will not show up if one only models a
24-hour average situation (as many traditional tools do).  Thus, we
suddenly are faced with a problem where we need to introduce more
dynamical correctness into the modeling while at the same time
considering much wider temporal and spatial scales than before.

It is fairly obvious that, when faced with a dynamical problem, a
``microscopic'' approach, i.e.\ starting with a description of the
smallest particles, is in terms of methodology the cleanest one.  In
transportation science, this currently means to consider individual
travelers rather than, say, aggregated link flows.  For example, it is
difficult to include individual route choice behavior into a
non-microscopic simulation.  There is also some agreement that the
currently most straightforward method to deal with microscopic
approaches in complicated real-world contexts is computer simulation,
as opposed to analytical techniques.  Now, when faced with a
compute-intense problem, such as systematic scenario evaluations (see,
e.g.,~\cite{Rickert:Wagner:Gawron,Beckman:etc:case:study,DYNASMART}), or
the simulation of the whole national transportation
system~\cite{nts:concept}, a very detailed and realistic
microsimulation (see, e.g.,~\cite{Wiedemann:model,Nagel:etc:flow-char})
may be computationally too slow, or too data-intensive to run.

Alternatives here are simplified models which still capture the
essentials of the dynamics at the transition to the congested regime.
Since traffic in general is dominated by the bottlenecks in the
system, these simulations concentrate on exactly these bottlenecks.
The most important bottlenecks in urban systems are traffic lights.
The natural outcome of this way of thinking are queuing-type
models~\cite{Simao:queueing,Gawron:simple}.  For vehicles that
enter the link, one calculates when they could arrive at the end of
the link.  When that time is reached in the simulation, they are added
to a queue at the end of the link.  They leave the queue once they
have advanced to its beginning.  The queue may have a limited service
rate, which models capacity restrictions.

This paper approaches this problem from a slightly different angle.
We use a very simple single-lane microsimulation to capture at least
some of the dynamics that is going on on the link itself.  This paper
will provide a systematic approach to such a model.  Sec.~2 will
describe our model, the way capacity restrictions are modeled, what
their behavior is, and what that means for the relation between the
simulation and reality.  In fact, capacity restrictions are simply
modeled by ``impurity sites'' or temporary ``blockages''
(e.g.~\cite{Janowsky:Lebowitz:blockage}).  Sec.~3 discusses an
implementation and some results for a Dallas study.  This is followed by a
short discussion, highlighting the differences between our approach
and other ``queuing-type'' approaches (Sec.~4), and a summary.

\section{A simplified approach}

We present a simple simulation model of city traffic, using a
combination of stochastic cellular automata (CA) and probabilistic
transitions between streets.  To represent the city network, we use
the usual definition (e.g.~\cite{Rickert:Nagel:DFW}) for links and
nodes: a link is a directed street segment, such as a bi-directional
road divided into two links, whereas a node is an intersection; a link
can also be defined by an input node and an output node.  Vehicles are
moved on a simple single-lane CA link, and are transferred from link to
link following a simple probabilistic law based on the link's
capacity.

\subsection{Links}

Links have different characteristics including length, speed-limit,
number of lanes, maximum capacity, etc.  The length is necessary to
adjust the number of sites needed for the discrete approach of the
CA. We use the standard reference of 7.5 meters for the length of one
site~\cite{Rickert:Nagel:DFW,Nagel:etc:flow-char,Nagel:Schreckenberg}.
Each site can be empty, or occupied by a vehicle with an integer
velocity $v \in \{0 \ldots v_{max}\}$. $v_{max} = 5$ gives good
agreement with physical experiments.

Since each link is considered as a one-lane segment, vehicles are
moved using the Nagel/Schreckenberg CA
rule~\cite{Nagel:Schreckenberg}. Summarizing the one-lane CA model,
the variable $gap$ gives the number of unoccupied sites in front of a
vehicle.  $p_{noise}$ is the probability to randomly be slower than
you could, and $rand$ is a random number between 0 and 1.  One
iteration consists of the following three sequential steps which are
applied in parallel to all cars:
\begin{enumerate}
\item
 Acceleration of free vehicles:
IF $(v < v_{max} )$ THEN $v=v+1$ 
\item
Slowing down due to other cars:
IF $(v > gap)$ THEN $v = gap$
\item
Stochastic driver behavior:
IF $(v > 0)$ AND $ ($ rand $< p_{noise}$) THEN $v=v-1$ 
\end{enumerate}

For each link, we introduce an intrinsic probabilistic transition,
which is a function of the capacity (maximum throughput).  The
one-lane model is faster and easier to implement compared to the
multi-lane CA.

%The introduction of a speed-limit will be optionnal in the simulation.

\subsection{Probabilistic transitions}

We introduce various probabilistic models to differentiate the
existing links within a city, from high capacity segments such as
freeways to low capacity segments such as arterials.  If we consider
only one-lane links, the probabilistic transition is introduced to
control the output flow of a link.  A high capacity link will produce
a high output flow, while a low capacity link will produce a low
output flow.

\subsubsection{Random traffic light}

\begin{figure}[t]
\centerline{\hbox{
\psfig{figure=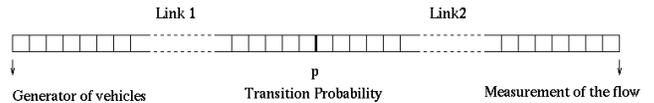,width=3.6in}}}
\caption{Schema of the experiment}
\label{fig:shema}
\end{figure}

\begin{figure}[t]
\centerline{\hbox{
\psfig{figure=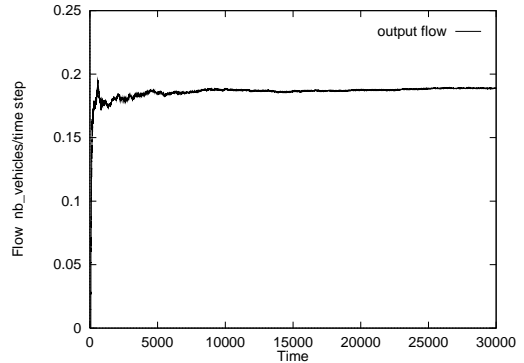,width=2.7in}}}
\caption{Flow versus Time for a transition probability of 0.5}
\label{fig:random0}
\end{figure}

\begin{figure}[t]
\centerline{\hbox{
\psfig{figure=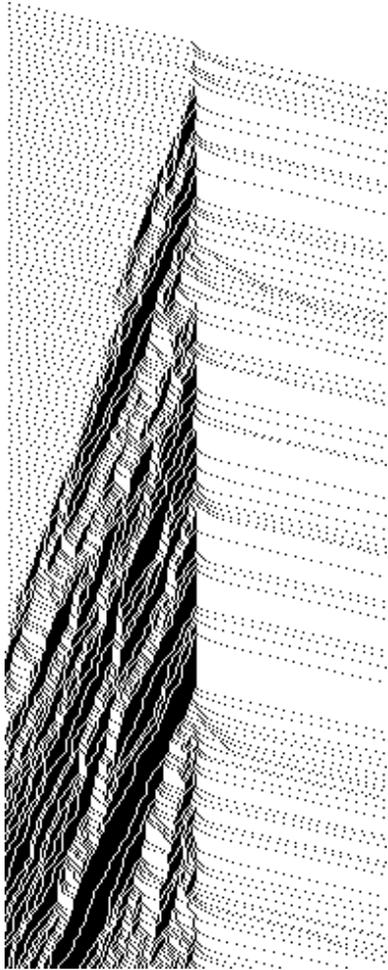,width=2.0in}}}
\caption{Space-Time diagram for a transition probability of 0.5}
\label{fig:random1}
\end{figure}

\begin{figure}[t]
\centerline{\hbox{
\psfig{figure=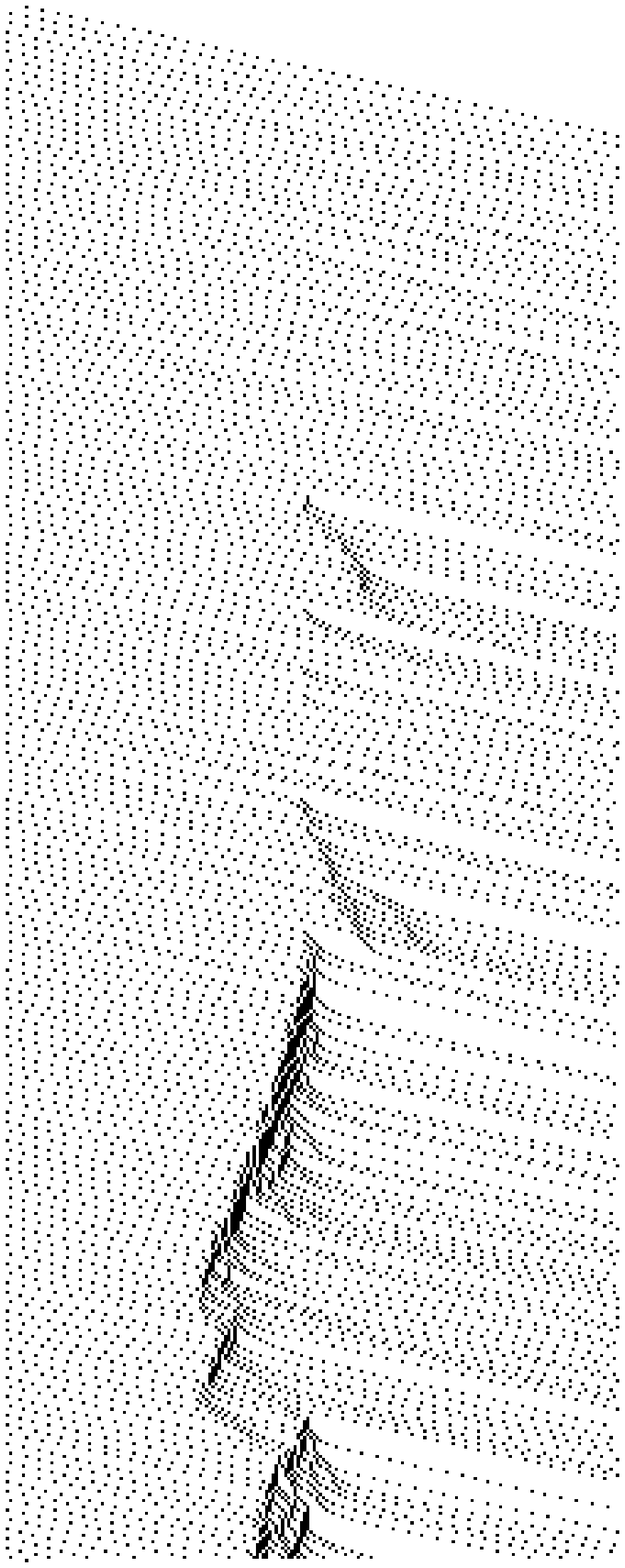,width=2.0in}}}
\caption{Space time diagram for p=0.9}
\label{fig:random3}
\end{figure}

\begin{figure}[t]
\centerline{\hbox{
\psfig{figure=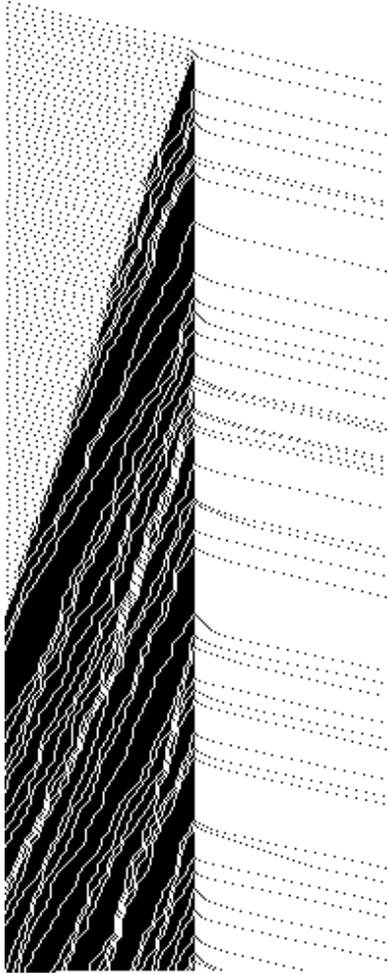,width=2.0in}}}
\caption{Space time diagram for p=0.2}
\label{fig:random4}
\end{figure}

Let us consider the experiment in fig[\ref{fig:shema}], consisting of
two consecutive links separated by a probabilistic transition
$p_{trans}$. The first site of link $1$ operates as a generator of
vehicles, where one vehicle is introduced per $n$ iteration(s).  The
flow measured at the end of the second link versus the number of
iterations is shown in fig[\ref{fig:random0}].  The probabilistic
transition is set to $0.5$ in this example.  The flow measured at
iteration $t$ is the number of vehicles that left the second link
until that moment, divided by $t$.  As a result, the unity of the flow
is vehicle per iteration.  We introduce 1 vehicle every 3~iterations
at the first site of the first link with maximum velocity 5. This is
enough to assure that the first link will reach around 1200 veh/h for
a $p_{noise}$ of 0.5, which is close to the maximum throughput of such
a link in the CA implementation~\cite{Nagel:Schreckenberg}.  If the
first site is not empty at the introducing time step, we do not add
vehicles.  The vehicle's velocities are updated by the one-lane CA
model before reaching the intersection.  If the vehicle is allowed to
go through the intersection by the CA forward rule, we check the
probabilistic transition.

If the generated random number is lower than the probability
$p_{trans}$, the vehicle keeps its velocity and reaches the second
link. In contrast, if the random number is greater than $p_{trans}$,
we place a fictitious car in the first site of the second link in
order to force the vehicle to brake and stop at the intersection.
Technically: If a car reaches the last five sites of a link, it
produces a random number.  We introduce the simple algorithm:
\begin{enumerate}
\item
Transition check:\\ IF $(rand<p_{trans})$ THEN normal CA-update ELSE
gap=distance from the vehicle to the intersection
\end{enumerate} 
This situation is in principle well understood.  The ``impurity site''
will create a reduced flow that can pass that site, and since flow
needs to be conserved along the link, this sets the maximum throughput
for the
link~\cite{Janowsky:Lebowitz:blockage,Yukawa:etc:blockage,Csahok:Vicsek:blockage,Vilar,Chung:Hui:blockage}.
Yet, in the context of the stochastic traffic cellular automaton as
used here, we are only aware of Ref.~\cite{Emmerich:Rank:blockage},
and the specific mechanism used there is not the one we wanted to use.
 
Figs [\ref{fig:random1}-\ref{fig:random3}] demonstrate the formation
of traffic jams spreading to the beginning of the link, caused by
braking of vehicles. The beginning of the second link can again be
considered as a generator of vehicles.  Nevertheless, the input flow
and $p$ are not proportional.

To illustrate this comment, we conduct the same experiment with
probabilistic transitions ranging from 0 to 1. The average flow obtained
for each experiment is presented in fig[\ref{fig:random2}]. For each data
point, the flow is averaged in the time period (5000,20000). See
fig[\ref{fig:random0}].  The intersection does not function as a perfect
generator of service rate $p$. If a vehicle leaves the last site of the
first link, this vehicle is not automatically replaced, due to the
stochastic third step included in the one-lane CA model. This plot can be
divided into three different parts:

% la vraie raison c'est que certain vehicules ne s'arrete pas a l'intersection

\begin{figure}[t]
\centerline{\hbox{
\psfig{figure=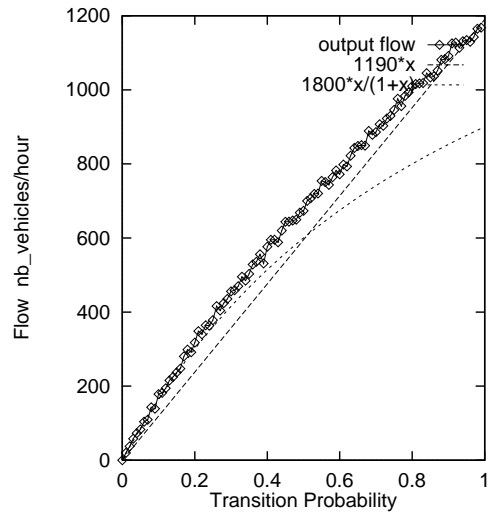,width=2.7in}}}
\caption{Flow versus transition probabilities}
\label{fig:random2}
\end{figure}

(i)~A high probabilistic transition ($p_{trans}$ between 0.8 and 1.0)
gives linear results with input flow. In this scenario, vehicles do
not stop often at the intersection, thus the intersection does not
work like a stop and start point.  See fig[\ref{fig:random3}].

(ii)~A low probabilistic transition (between 0 and 0.4) gives results
that can be explained by a simple hypothesis.  Most of the cars stop
at the intersection and form a compact traffic jam, as shown in
fig[\ref{fig:random4}]. There are no important spaces in this queue.
Assuming that the second last site of the first link is always
crowded, how many iterations does a vehicle need to go through the
intersection?  If a vehicle is on the last site of link $1$, the
vehicle needs $1/p_{noise}$ iterations on average to advance, and then
multiplied by $1/p_{trans}$ to go through the intersection. Viewed
from the perspective of the next following vehicle, that one needs to
wait $1/p_{noise} \cdot 1/p_{trans}$ steps until the vehicle ahead is
gone, and then another $1/p_{noise}$ steps to move itself to arrive at
the last site.  As a result the average number of iterations for a
vehicle to advance from the second last site of link $1$, to the
intersection is $1/p_{noise} + 1/(p_{noise} p_{trans})$.  This could
in theory be continued, but it would not necessarily get better
because one would need to include the influence of ``holes'' in the
queue; or, more technically: The approximation is only valid for
$p_{trans} \to 0$, and second order corrections are thus negligible.
In any case, the corresponding flow is $F \approx p_{noise} p_{trans}
/ (1+p_{trans})$.
 
The function, $F$, shown in fig[\ref{fig:random2}] fits well to the
data measured for low values of $p$, while for $p \geq 0.4$ the
hypothesis is no longer valid.

(iii)~Figure [\ref{fig:random1}] demonstrates what happens for
probabilistic transitions between 0.4 and 0.8 at a microscopic level.
Within the queue, holes are generated by the intersection and an
analytical approach becomes more difficult. Periodically, vehicles
pass through the intersection without braking and stopping, which
produces a higher flow compared to the linear relationship illustrated
in fig[\ref{fig:random2}].

\begin{figure}[t]
\centerline{\hbox{
\psfig{figure=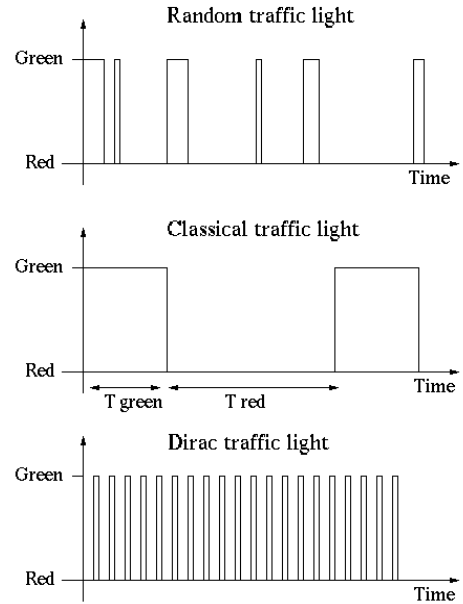,width=2.8in}}}
\caption{Different distribution probabilities}
\label{fig:distribution}
\end{figure}

Many experiments can be conducted using other probability
distributions for the intersection. The model previously described
operates like a random traffic light, where the light becomes green
with the probability $p_{trans}$, which is also the fraction of the
time the light is green: $f_{green} = p_{trans}$.  This model can be
considered to be one between two extreme distributions, where in
between the extreme cases one can encounter an infinite number of
distributions that keep the fraction of a green light of the total
time of a traffic cycle constant.  The first distribution is a
classical traffic light. The green fraction here is straightforward:
$f_{green}=T_{green}/(T_{green}+T_{red})$. We call the second model a
Dirac traffic light. As we work with discrete systems, the objective
is to set a green light or a red light on only one time unit, equally
spaced on a cycle. The green fraction is $f_{green} = 1/(1+T_{red})$
for $T_{red} \ge 1$ (and $T_{green} = 1$ by definition) or
$1-1/(1+T_{green})$ for $T_{green} \ge 1$.  All three distributions
are illustrated in fig [\ref{fig:distribution}].
 
Next, we present the same experiments discussed above, for these two distributions.

\subsubsection{Normal traffic light}

We repeat the same experiment described in fig[\ref{fig:shema}]
with a normal traffic light at the intersection.

\begin{figure}[t]
\centerline{\hbox{
\psfig{figure=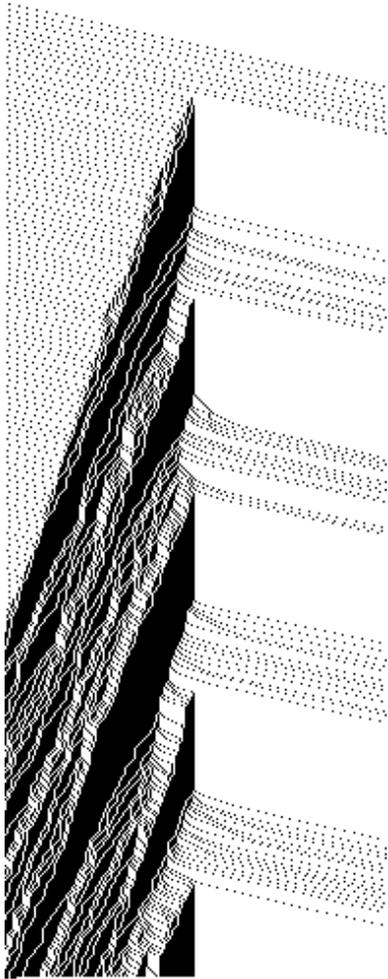,width=2.0in}}}
\caption{Space-time diagram normal traffic light p=0.5}
\label{fig:normal1}
\end{figure}

\begin{figure}[t]
\centerline{\hbox{
\psfig{figure=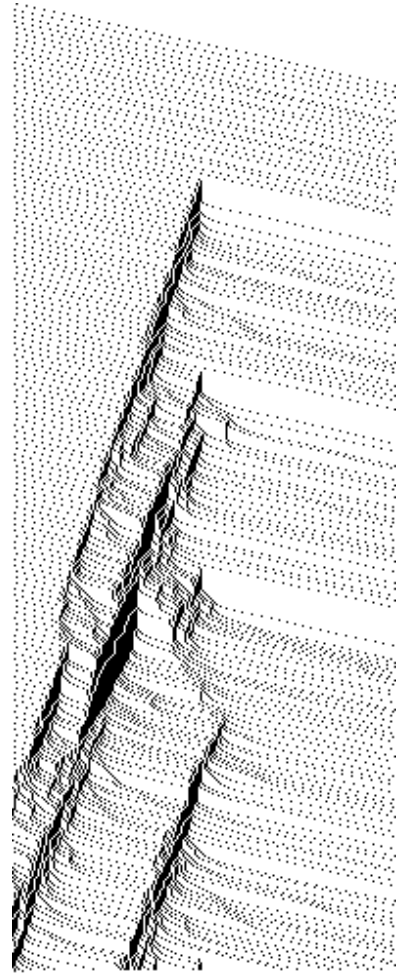,width=2.0in}}}
\caption{Space-time diagram normal traffic light p=0.9}
\label{fig:normal2}
\end{figure}

\begin{figure}[t]
\centerline{\hbox{
\psfig{figure=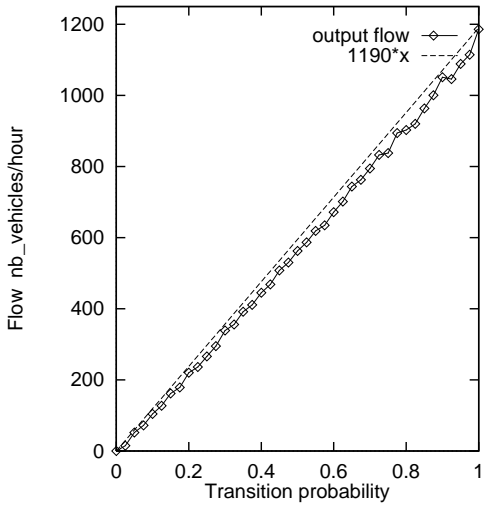,width=2.7in}}}
\caption{Flow versus Transition probability}
\label{fig:normal4}
\end{figure}

The dissolution of a queue as the light turns periodically green is shown
in fig[\ref{fig:normal1}].  This phenomenon does not provide an easy
analytical solution.  For each green fraction~$f_{green}$ ranging from 0 to
1, the input flow of the second link is measured and is illustrated in
fig[\ref{fig:normal4}].  This relationship is almost linear. For high
values of transitions, vehicles still have to stop occasionally, which
decreases the output flow. Figure[\ref{fig:normal2}], when compared to the
space-time diagram produced by the random traffic light
fig[\ref{fig:random3}], displays a lack of fluidity.

\subsubsection{Dirac traffic light}

\begin{figure}[t]
\centerline{\hbox{
\psfig{figure=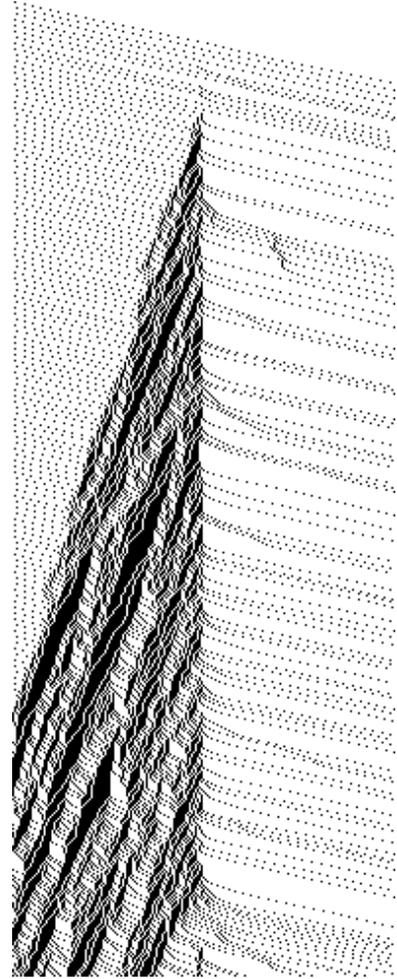,width=2.0in}}}
\caption{Space-time diagram Dirac traffic light p=0.5}
\label{fig:Dirac0}
\end{figure}

The Dirac traffic light generates the highest flow for a given
$f_{green}$ in the experiment or fig[\ref{fig:shema}]. The space-time
diagram performed with a probabilistic transition of 0.5 is given as
an example.  In this case, the traffic light is successively green and
red.  Figure [\ref{fig:Dirac0}] shows less compact traffic jams at the
end of the first link than the other space-time diagrams for the same
probabilistic transition.  This is still due to the vehicles that pass
through the intersection at maximum velocity without braking.  The
analytic explanation for this is the fact that the parallel update
tends to generate states where particles are followed by holes,
sometimes called ``particle-hole
attraction''~\cite{Schreckenberg:etc:Ito}.

\begin{figure}[t]
\centerline{\hbox{
\psfig{figure=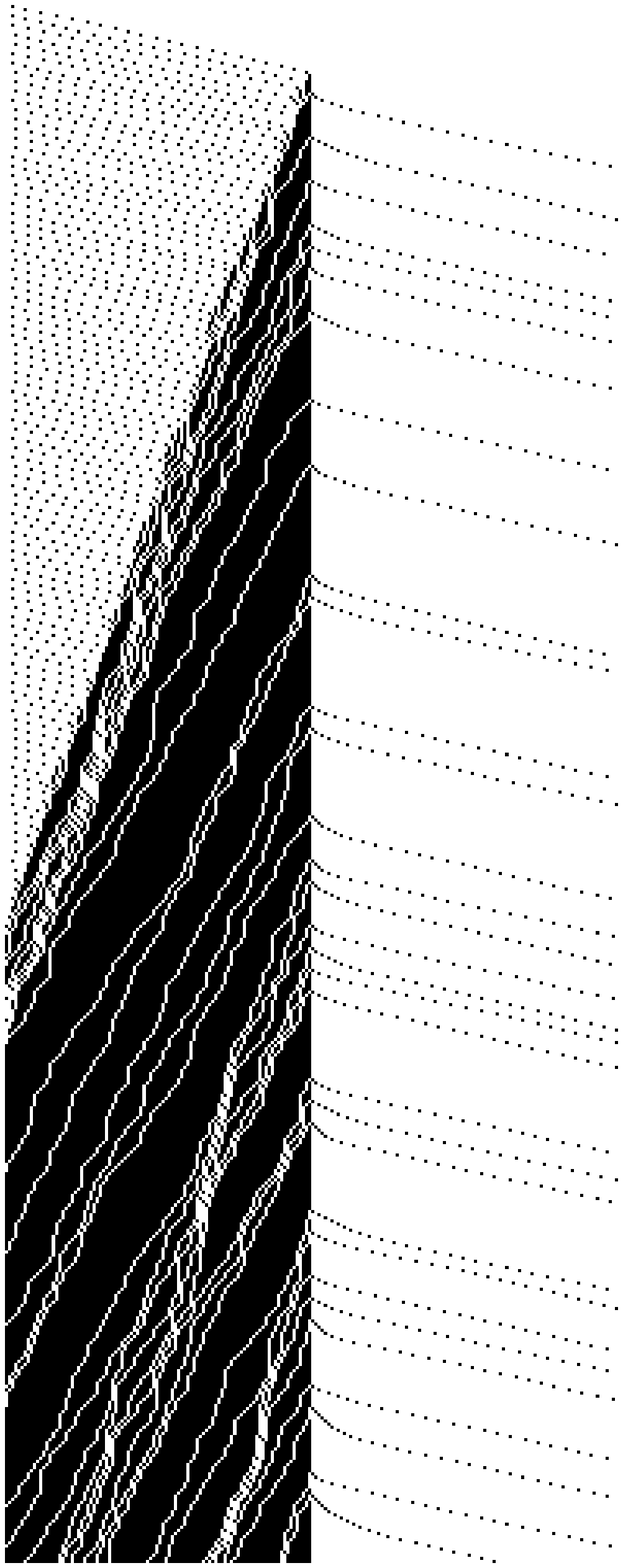,width=2.0in}}}
\caption{Space-time diagram Dirac traffic light p=0.16}
\label{fig:Dirac2}
\end{figure}

\begin{figure}[t]
\centerline{\hbox{
\psfig{figure=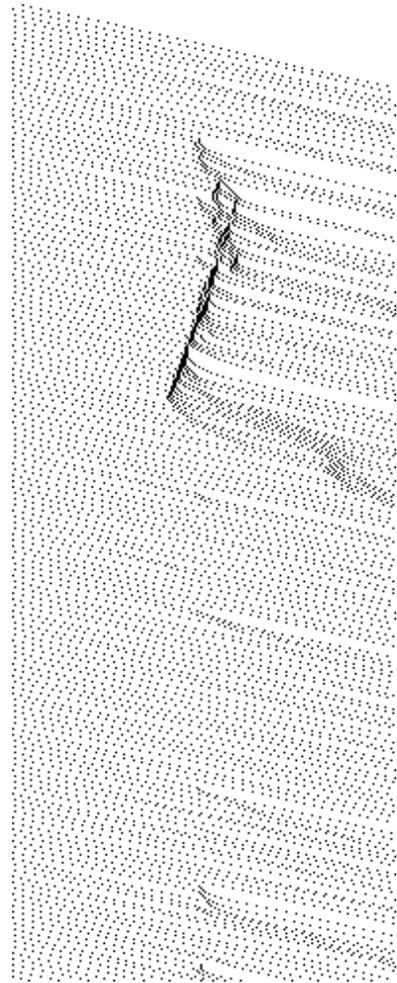,width=2.0in}}}
\caption{Space-time diagram Dirac traffic light p=0.9}
\label{fig:Dirac5}
\end{figure}

The output flow of link $1$ for any value of $p$ is much higher
than the two flows measured previously for the two other probability 
distributions. There is no linear relation at any position on this diagram.
The space-time diagrams plotted for a $p=0.16$ and $p=0.9$
exhibit more fluidity for the output traffic fig[\ref{fig:Dirac2},\ref{fig:Dirac5}].

\begin{figure}[t]
\centerline{\hbox{
\psfig{figure=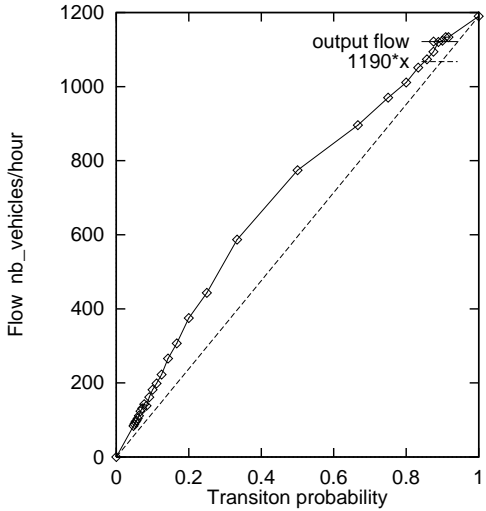,width=2.7in}}}
\caption{Flow versus transition probability}
\label{fig:Dirac4}
\end{figure}

\section{Dallas}

\subsection{Implementation}

The normal traffic light model is the most linear model simulated in
this paper.  On the other hand, setting a traffic light at each
intersection would cost computation time.  The random traffic light
presents the advantage to be checked only when a vehicle reaches the
intersection. The vehicle generates a random number which allows it to
drive trough the intersection or not.  

We apply this model to the Dallas Fort-Worth area.  The context is the
so-called Dallas/Fort Worth case
study~\cite{Beckman:etc:case:study,Nagel:Barrett:feedback}
 which has been done as part of the TRANSIMS
(TRansportation ANalysis and SIMulation System)
project~\cite{TRANSIMS}.  TRANSIMS uses individual route plans for
each individual traveler.  A route plan consists of a starting time, a
starting location, a list of links the vehicle intends to follow, and
an ending location.  A microsimulation in the TRANSIMS project such as
the one described here is thus faced with the task to move these
vehicles according to these specifications.

One immediately observes that one somehow has to correct for the fact
that we are only using single-lane roads, i.e.\ our links will usually
not be able to carry the prescribed number of vehicles.  We solve that
problem by using a sub-sample of the plans.  The size of that
sub-sample is obtained as follows:\begin{itemize}

\item
$p_{noise}=0.5$ results in a maximum throughput of a link of
approximately~$1200$~veh/h (using $p_{trans}=1$).

\item
We search for the link with the highest capacity in the area we want
to simulate.  In our case, this was a four lane freeway with a
capacity of~7800~veh/h.

\item
We thus need to sub-sample the population by a factor of $1200/7800 \approx
0.154$, i.e.\ a route plan from the full plan-set is going to be used with a
probability of 0.154.  

\item
Links which have a lower capacity than 7800~veh/h take this fact into
account by using a value of $p_{trans}$ according to
Fig.~\ref{fig:random2}, i.e.\ if the value of the road is $C$, then
the value $C \cdot 0.154$ is used on the y--axis to find the correct
value of $p_{trans}$ on the x--axis.

\end{itemize}
A more precise calibration is more complicated than this because it
also depends on the interplay between route planning and route
execution.  This is clearly out of the scope of this paper; further
publications on the subject are in preparation.

\subsection{Simulation results}

In this section, we want to give some examples how this simulation is
going to be used.  These examples will be given in the context of the
TRANSIMS Dallas/Fort Worth case study.  That case study used as input
a street network of the Dallas/Fort Worth area, containing 24662 links
and 9864 nodes, and information of all trips in this area during a
24~hour period (approx.\ 10~million trips).  The study focused on a busy
5~miles times 5~miles area north of downtown Dallas, and on the time
between 5am and 10am.  This still involved 300\,000~trips.  As
mentioned above, micro-simulations in the TRANSIMS project are
route-plan driven.  Thus, for each of these 300\,000~trips, route
plans were calculated.  The fact that drivers adjust to congestion
caused by other drivers was taken into account by iteration several
times between the route planning and the micro-simulation.  For
further information, see
Refs.~\cite{Nagel:Barrett:feedback,Rickert:feedback,Nagel:etc:comparisons}.

One important specification missing in the above description of the
micro-simulation is how vehicles enter and leave the simulation.
TRANSIMS specifies parking locations along links, which represent all
parking opportunities that can be reached from this link.  In order to
prevent that the traffic that leaves parking unrealistically disturbs
the traffic flow, vehicles from the parking locations are only
inserted if $v_{max}$~sites backwards from the parking location are
empty.  If the space is not free, the car
is placed in a queue, waiting to enter the simulation in one of the
following iterations.

\begin{figure}[t]
\centerline{\hbox{
\psfig{figure=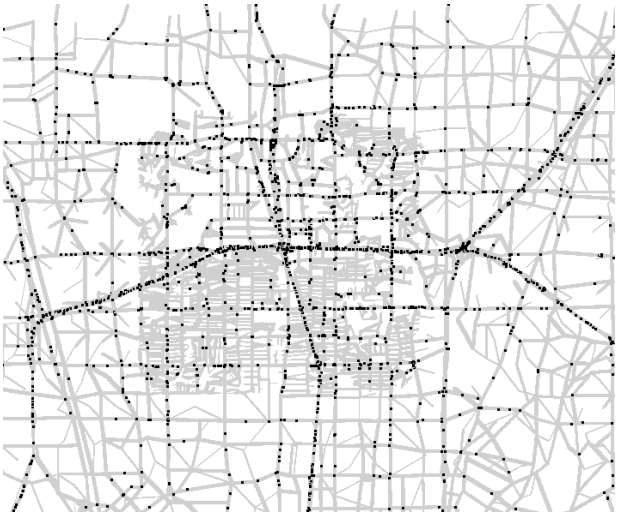,width=3.2in}}}
\caption{Snapshot of the case study at 7:00}
\label{fig:instantane}
\end{figure}

A snapshot of such a simulation with the model described in this paper
can be found in Fig[\ref{fig:instantane}].  The denser square area in
the center represents the study area, where all streets including
local streets were represented in the data base.  For this example,
also the streets outside that area were simulated.  Dots denote
individual vehicles.  In this plot, most of the traffic is on the
freeways, as is realistic.  Also, one notes that for lower capacity
road, traffic is mostly queued up towards the end, as one would expect
from the dynamics of the model.  Yet, this is really not too
unrealistic since also in reality traffic through minor roads tends to
queue up at the ends.

% link ids
% 519350302
% 5193602
% 5193701
% 519400602
% 519400601

\begin{figure}[t]
\centerline{\hbox{
\psfig{figure=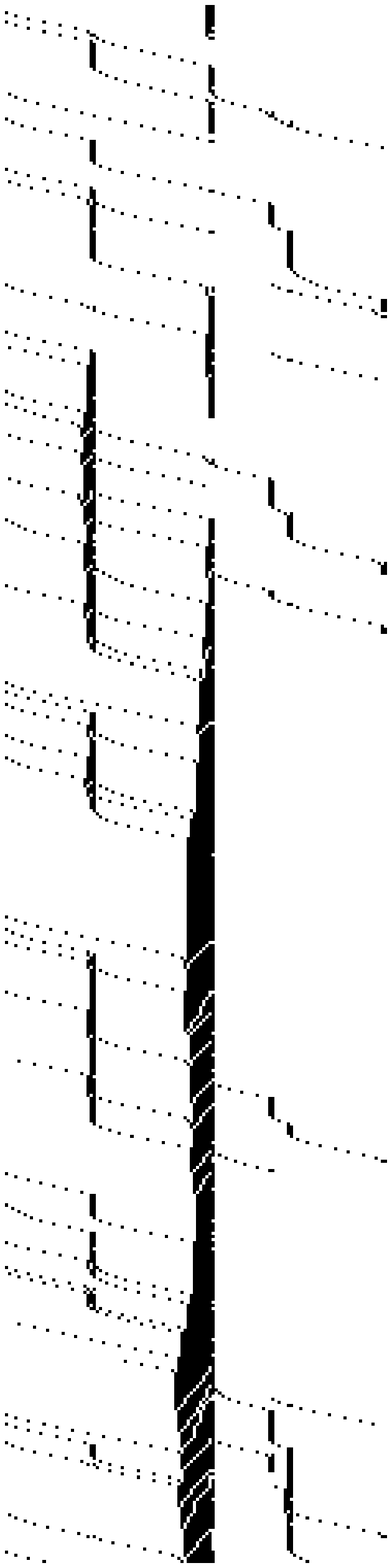,width=1.0in},
\psfig{figure=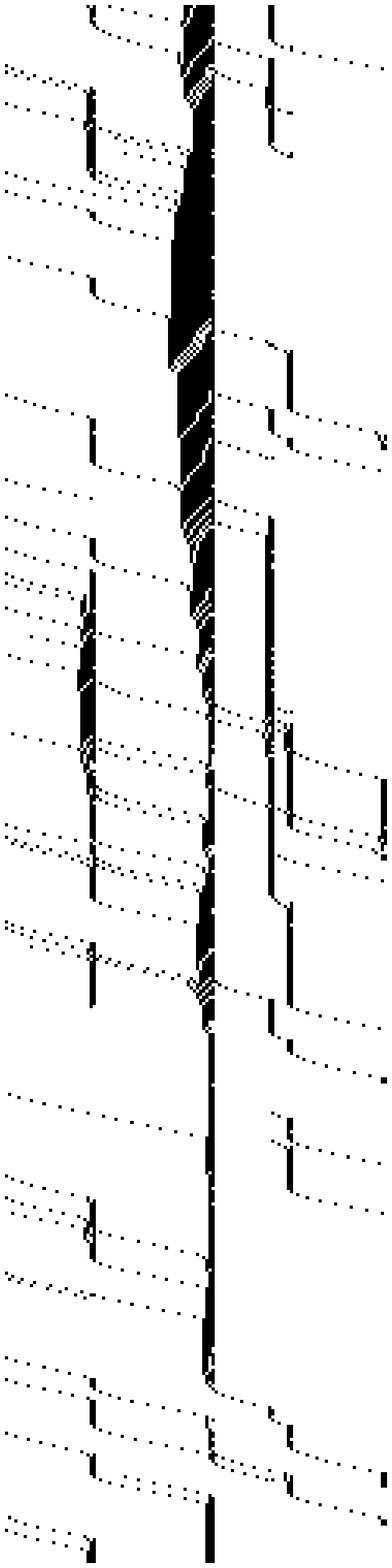,width=1.0in},
\psfig{figure=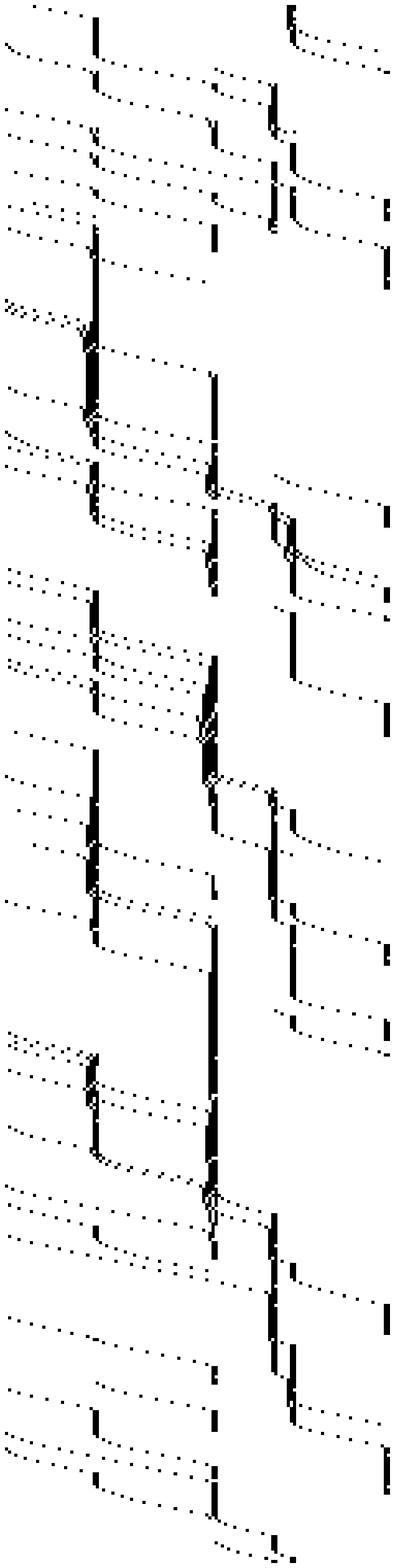,width=1.0in}
}}
\caption{Space-time plot of a particular link (Beltline Rd., an east-west
arterial in the northern part of the area) from 7:00am to 7:05am (left),
7:30am to 7:35am (middle), and 8:00am to 8:05am (right.)}
\label{fig:space_time}
\end{figure}

The space-time diagram of five consecutive links is shown in
fig[\ref{fig:space_time}].  These links are a part of an east-west
arterial located in the north of the study area.  The figure shows
nicely how queues built up at the end of links due to the capacity
restrictions.  

\begin{figure}[t]
\centerline{\hbox{
\psfig{figure=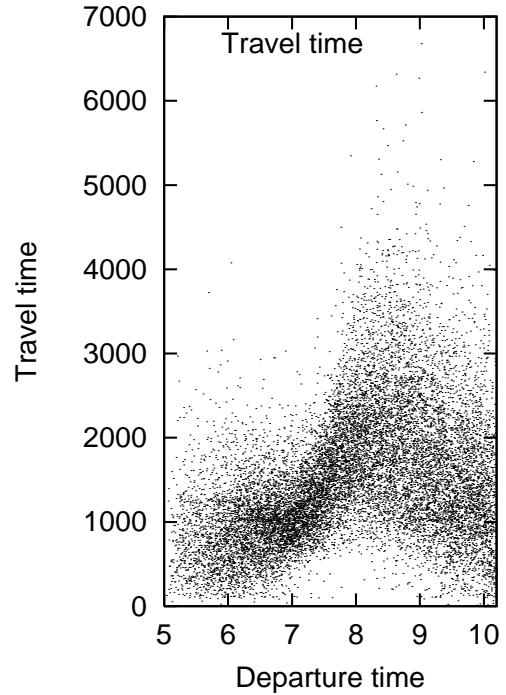,width=2.7in}}}
\caption{Travel times versus departure time}
\label{fig:ttimes}
\end{figure}

Many statistics can be extracted from the simulation.  As a further
example, we present the travel time versus departure time for each
vehicle (Fig.~\ref{fig:ttimes}).  This figure shows nicely that even
such a simple simulation as the one described in this paper can, given
a realistic trip demand input, display the higher travel times during
the rush hour.

\subsection{Computational performance}

\begin{figure}[t]
\centerline{\hbox{
\psfig{figure=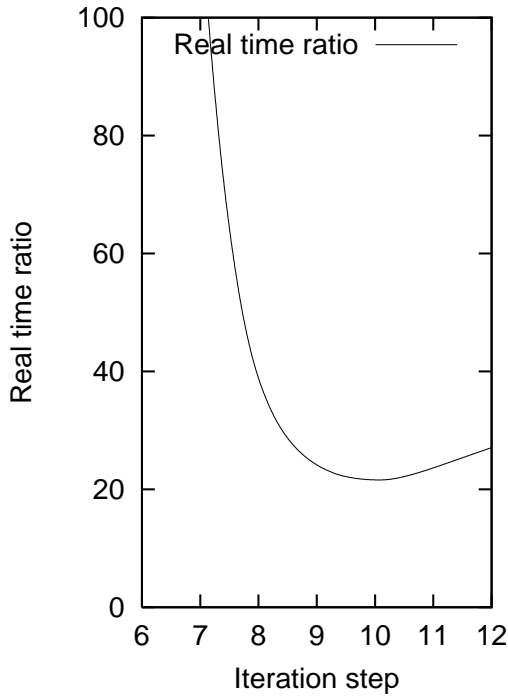,width=2.7in}}}
\caption{Real time ratio}
\label{fig:realtime}
\end{figure}

We present a performance diagram in fig[\ref{fig:realtime}] where we
introduce the RTR versus the simulation time.  The RTR is the ratio of
the real time on the simulation time. This example of simulation was
executed on a SUN UltraSparc CPU with 250 MHz where approximately
46000~plans were simulated in the whole Dallas Fort-Worth area.  The
diagram fig[\ref{fig:realtime}] shows a ratio of 23 in the middle of
the rush period, but in average the ratio is around 28.  This clearly
shows that simulations like the one described here have enough
computational speed for thorough investigations of traffic problems.

\section{Discussion}

Transportation models using simplified link dynamics ultimately fail
to generate some aspects of a complicated reality, such as a turn
pocket having a queue spill-back into the lanes going straight.  Yet,
as pointed out in the introduction, using a highly realistic model
sometimes is not an option, for example because of computational
constraints or data collection constraints.  In such cases, knowing
the different limitations of the simplified models becomes crucial.

For example, the different simplified models handle queue discharge
dynamics in different ways.  In our model, when a vehicle is moved to
the next link, this leaves a ``hole'' on the link where the vehicle
comes from.  In the next time step, that hole may or may not be filled
by an advancing vehicle, according to the stochastic driving rules.
In a congested situation, this hole slowly travels backwards, until it
eventually reaches the other end of the link, allowing a new vehicle
to enter the link.  In Ref.~\cite{Gawron:simple}, holes are
transmitted instantaneously to the other end of the link.  The method
of Ref.~\cite{Simao:queueing} assumes infinite queuing capacity on
each link.  It is clear that all three methods will generate different
dynamics.

As another example, in fairly realistic models, sources and sinks for
traffic are better located in the middle of links instead of at nodes.
Vehicles attempt to squeeze in between other vehicles at that location
on the link.  If the link gets congested, the additional vehicles will
have trouble finding additional space to squeeze in.  It is clear that
models who totally give up a representation of traffic dynamics on the
link such as~\cite{Gawron:simple,Simao:queueing} will lead
to different behavior for traffic sources and sinks.

Certainly, the simple queuing models could compensate for that.  Yet,
that usually comes at the price of being tedious.  Often, it will be
more straightforward to move directly to a higher fidelity (but
usually computationally slower and more data intensive)
micro-simulation.  We believe that, at the current stage, it is more
important to really understand the dynamical differences between
different models and to compare their behavior in real-world
applications, than to attempt to improve simplified models in
non-intuitive ways.

Last but not least, the model presented in this paper actively moves
vehicles along links with roughly realistic dynamics.  This makes
graphical output such as in Fig.~\ref{fig:instantane} much more
intuitive and appealing.

\section{Summary}

``Blockage'' sites, i.e.\ sites which move particles or vehicles only
a fraction of the time, reduce the maximum throughput of a link of
cellular automata models for traffic flow and particle movement
studies.  We have systematically tested the effects of three different
blockage schemes, where one was the usual random draw, one was a
regular traffic light with long red and green times, and one was what
we called a ``Dirac'' traffic light because it had 1-second spikes of
one color.  In general, there is no linear relation between the
fraction of green time and the throughput.  The Dirac traffic light
returned the highest throughput; the explanation for this is the
``particle-hole'' attraction that can be found in the type of cellular
automaton that was used.  Since none of the timing schemes returns a
totally linear relation, we used the random scheme in an
implementation of traffic in Dallas.  We showed some exemplary results
of this implementation.

\section{Acknowledgments}

This work has been performed at Los Alamos National Laboratory, which
is operated by the University of California for the U.S. Department of
Energy under contract W-7405-ENG-36.

\bibliographystyle{prabib} 
\bibliography{ref,kai}
\end{document}